# Dynamics of electrochemical flows II: Electrochemical flows-through porous electrode

Chengjun Xu[1*], Chin-Tsau Hsu[1,2]

[1]Graduate School at Shenzhen, Tsinghua University, Shenzhen City, Guangdong Province, 518055, China. E-mail address: vivaxuchengjun@163.com.

[2]Department of Mechanical Engineering, Hong Kong University of Science and Technology, Kowloon, Hong Kong. E-mail address: mecthsu@ust.hk.

**Abstract:** The electrolyte (comprising of solute ions and solvents) flow-through the porous media is frequently encountered in nature or in many engineering applications, such as the electrochemical systems, manufacturing of composites, oil production, geothermal engineering, nuclear thermal disposal, soil pollution. In this study, we provide a new general theory for the electrochemical flows-through porous media. We use static method and set up two representative elementary volumes (REVs). One is the macroscopic REV of the mixture of the porous media and the electrolyte, while the other is the microscopic REV in the electrolyte fluid. The establishment of two REVs enables us to investigate the details of transports of mass, heat, electric flied, or momentum in the process of the electrochemical flows-through porous electrode. In this work, the macroscopic governing equations are derived from the conservation laws in the macroscopic REV to describe the electrochemical flows-through porous media. At first, we define the porosity by the volume and surface and divide the porous media into various categories. Then the superficial average is transformed into intrinsic averages to derive the interaction terms between the solid and the fluid, known as terms of dispersion, tortuosity and interfacial transfer. The macroscopic governing equations are derived by performing the intrinsic average on the microscopic governing equations. After done that, the unknown terms related to the dispersion, tortuosity and interfacial transfer are emerged in the governing equations.

1. Introduction

The fluid (liquid or gas) flow-through the porous media is frequently encountered in nature or in many engineering applications, such as the electrochemical systems, manufacturing of composites, oil production, geothermal engineering, nuclear thermal disposal, soil pollution, to name a few. Electrochemical systems, for example, electrochemical reactors, batteries, supercapacitors, fuel cells, etc, are widely used in modern society. The energy crisis put the electrochemical conversion and storage systems into the center of today's scientific research around the world. The electrochemical systems generally involve a common and fundamental process, in which the electrolyte (or gas for gas electrode) fluid flows through the porous electrode. The motion of the electrolyte in the electrochemical systems leads to the flows of solvent and ions, known as electrochemical flows. The porous electrode provides a very lager surface area, on which the electrochemical reactions and electrochemical double layer occur, to generate a large current. The improvement and optimum of the electrochemical systems rely on the mathematical modeling or general understanding of the mechanism occurring in such systems. During the past decades J. Newman has done excellent works on the porous electrode theory[1-3]. He made the first breakthrough work on the flows-through porous electrode by disregarding the actual geometric detail of the pores in the porous electrode from the macroscopic point of view. Newman's porous electrode theory explains the mass distribution in the electrochemical systems and current-voltage response very well. R.E. White has also done lots of works on the modeling of many electrochemical systems, for example lithium ion and nickel cadmium batteries, based on the isothermal electrochemical model, which is derived from the Newman's porous electrode theory by many researchers [4-6].

The future development of the electrochemical systems needs the general clarification of the details of the transport phenomena at the interface between the solid and the electrolyte fluid, for example the transference of the mass, heat and electric force and the velocity of the ions and solvents, in the porous electrode. These details enable the electrochemical systems to work more durable, safe, reliable, and workable with a better performance. For example, the improvement of the safety issue of lithium ion batteries is dependent on the general understanding of thermal runaway in the porous structure in detail. In addition, the fundamental equations occurring in the natural and physical sciences have to be obtained from conservation laws. Conservation laws are just balance laws, with equations to express the balance of some quantities throughout a process. Mathematically, conservation laws usually translate into integral/differential equations, which are then regarded as the governing equations of the process. Therefore the fundamental equations governing the electrochemical flows-through the porous electrode have to be derived from the conservation laws.

In our work, we try to establish the general theory for the electrochemical flows-through porous media from the basic conservation laws of charge, mass, momentum, energy and mass concentration. We use static method and set up two representative elementary volumes (REVs).

One is the macroscopic REV of the mixture of the porous media and the electrolyte, while the other is the microscopic REV in the electrolyte fluid. Figure 1 shows the schematic picture of microscopic and macroscopic REVs in the domain of electrolyte flowing through porous electrode. The establishment of two REVs enables us to investigate the details of transports of mass, heat, electric flied, or momentum in the process of the electrochemical flows-through porous electrode. In our previous paper, we derived the governing equations from the conservation laws of charge, mass, momentum, energy and mass concentration in the microscopic REV to describe the electrochemical flows [7]. In this paper, we investigate the conservation of charge, mass, momentum, energy and mass concentration in the macroscopic REV. The governing equations for the electrochemical flows-through porous electrode will be derived from the conservation laws in mixture of the electrolyte fluid and the porous media. At first, we define the porosity from volume and area and divide the porous media into various categories. Then we transform the superficial average into intrinsic average to get the fluid-solid interfacial interaction terms. The macroscopic governing equations are derived by performing the intrinsic volume average on the microscopic equations. After done that, the terms of dispersion, tortuosity, and solid-fluid interfacial transfer of mass, heat, electric field, potential and electric force are emerged in the macroscopic equations as the unknown terms.

2 Definition of macroscopic representative elementary volume

The pressure created by the flows in porous media is written as the Darcy law,

$$-\nabla p = \frac{\mu U}{K} \qquad (1)$$

where $p$ is the pore pressure, $\mu$ is fluid viscosity, $U$ is the Darcy velocity. K is the permeability empirically determined as

$$K = \frac{\phi^3 d_p^2}{a(1-\phi)^2} \qquad (2)$$

where $\phi$ is the void porosity, $d_p$ is the particle size of solids, and $a$ is a non-dimensional constant usually taken as 180.

However, the Darcy law is more suitable for the fluid with low Reynolds number and misses the important details near the wall. To solve this issue, we can study the porous media by scaling of a macroscopic representative elementary volume (REV) larger than the scale of the pore structure. Therefore, the average process can be done for quantities in each phase without ambiguity.

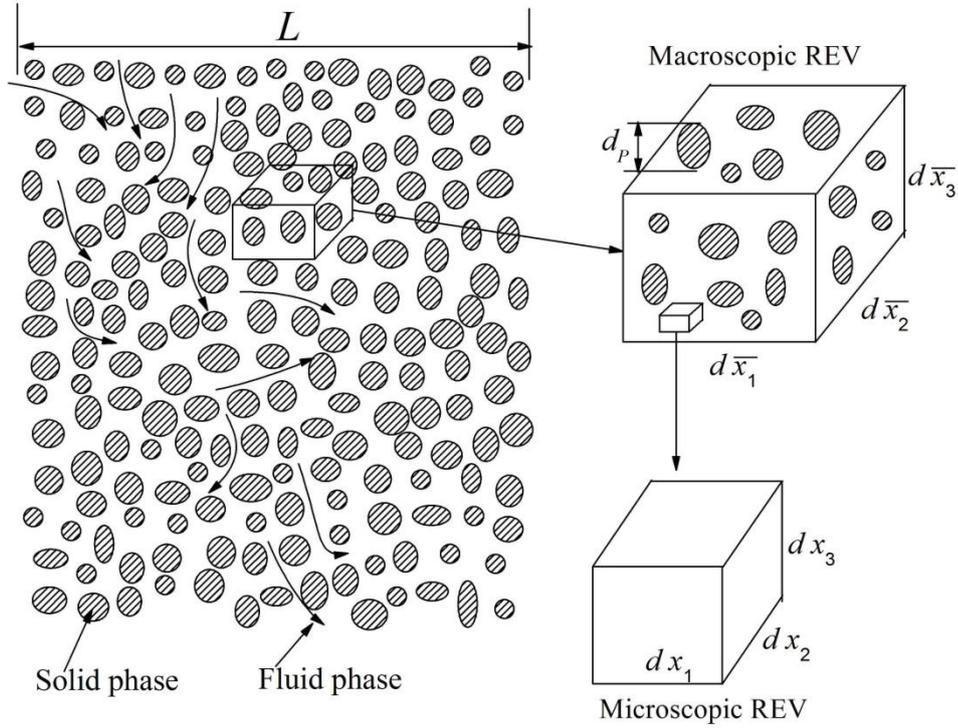

Figure 1 Schematic images of macroscopic and microscopic representative elementary volumes (REVs)

Figure 1 shows the definition of a macroscopic representative elementary volume. In the porous media with a length scale of $L$, we define a macroscopic REV (mREV) with a length scale of $l$. The length scale $l$ is much larger than the particle size of $d_p$, but much smaller than the length scale of $L$. Note that there is still a small microscopic REV (sREV) with the scale length of $dx_i$ in the fluid. The scale length of $dx_i$ is much smaller than the particle size of $d_p$, but much larger than the molecular scale of $l_m$. The scaling law is summarized in the Table 1.

Table 1 Scaling law

| Variable | Physical meaning | Variable | Physical meaning |
| --- | --- | --- | --- |
| $\bar{x}_i$ | Macroscopic length scale variable | $L$ | Global (macroscopic) length scale |
| $l$ | length scale of large REV | $d_p$ | particle size |
| REVs | $V = d\bar{x}_1 d\bar{x}_2 d\bar{x}_3$, $dV = dx_1 dx_2 dx_3$ | $x_i$ | The scale length of small REV |
| $l_m$ | Molecular length scale | | $L \gg l \gg d_p \gg dx_i \gg l_m$ |

The physical quantities can only be defined after scaling of mREV. The porosity ($\phi$) is defined in the mREV.

$$\phi = \frac{V - V_1}{V} = \frac{V_2}{V} \qquad (3)$$

where $V_1$ is the volume of the solid phase in the mREV, $V_2$ is the void volume (fluid volume) in the mREV and $V$ is the total volume of the mREV.

If the $\phi$ is independent of the size of $V$ chosen, we get

$$\phi = \frac{V_2}{V} = \frac{V_2 + \Delta V_2}{V + \Delta V} = \frac{\Delta V_2}{\Delta V} = \frac{A_2 dn}{A dn} = \frac{A_2}{A} = \phi_a \qquad (4)$$

where $\phi_a$ is the areal porosity. $A$ is the total surfaces enclosed by mREV. $A_1$ and $A_2$ are the solid-solid and fluid-fluid surfaces, respectively.

Here $A_2$ and $A$ refer to the area enclosing $V$. However, the definition for $\phi_a$ is generally referred to arbitrary large area $A_2$ and $A$. A formal definition of $\phi_a$ will be given later. The important point is that as in a Cartesian coordinate system, $\phi_a$ may depend on plane orientation.

Based on the definition of porosity, there are categories of porous media. When the porosity is constant everywhere it is homogeneous porous media. If the porosity changes with the position, it is inhomogeneous. If the area porosity is independent of plane area orientation, it is isotropic media. Anisotropy can be resulted from the anisotropy properties of solids and/or fluid. If the porosity does not change with the time, it is rigid media, otherwise it is deformable media. The expression of the porosity depends on the specific case and needs to be defined in advance.

3 Superficial averages

The physical quantities (**w**) can be defined after the definition of the mREV. There are intrinsic volumetric average and superficial volumetric average. The intrinsic average ($\overline{\mathbf{w}}$) is the average of a fluid quantity **w** over the fluid volume.

$$\overline{\mathbf{w}} = \frac{1}{V_2} \int_{V_2} \mathbf{w} \, dV \qquad (5)$$

Superficial average (W) is the average of a fluid quantity **w** over the total volume

$$W = \frac{1}{V}\int_{V_2} \mathbf{w} dV = \frac{V_2}{V}\frac{1}{V_2}\int_{V_2} \mathbf{w} dV = \phi\bar{\mathbf{w}} \qquad (6)$$

This equation establishes the relation between the intrinsic average and the superficial average. Note that when **w** is equal to 1, the superficial average (W) is equal to $\phi$. The intrinsic average has more physical meaning than superficial average.

The divergence theorem of the intrinsic average is

$$\int_{V_2} \nabla \cdot \mathbf{w} dV = \int_{A_2} \mathbf{w} \cdot d\mathbf{s} \qquad (7)$$

where $d\mathbf{s}$ is the surface area increment vector and is represented by $\mathbf{n}dA$.

The superficial average of divergence is

$$\frac{1}{V}\int_{V_2} \nabla \cdot \mathbf{w} dV = \frac{1}{V}\int_{A_2} \mathbf{w} \cdot d\mathbf{s} + \frac{1}{V}\int_{A_{12}} \mathbf{w} \cdot d\mathbf{s} = \bar{\nabla} \cdot \left(\frac{1}{A}\int_{A_2} \mathbf{w} d\mathbf{s}\right) + \frac{1}{V}\int_{A_{12}} \mathbf{w} \cdot d\mathbf{s} \qquad (8)$$

where $\bar{\nabla}$ is the macroscopic gradient operator, $A$ is overall area of mREV, $A_2$ is fluid-fluid surface area, and $A_{12}$ is the fluid-solid interfacial surface area.

The superficial average of divergence can be written as

$$\phi\overline{\nabla \cdot \mathbf{w}} = \bar{\nabla} \cdot (\phi\bar{\mathbf{w}}) + \frac{1}{V}\int_{A_{12}} \mathbf{w} \cdot d\mathbf{s} \qquad (9)$$

The last term represents the fluid-solid interfacial transfer. Note that by using the above definition the areal porosities in different directions are the same, identically equal to $\phi_a$. And areal and volumetric averages can be taken as the same in the domain of a randomly packed porous media. By applying the above divergence theorem on a scalar w(w**i**,w**j**,w**k**) to w**i**, w**j** and w**k**, respectively and summing up, the superficial average of gradients is

$$\phi\overline{\nabla w} = \bar{\nabla}(\phi\bar{w}) + \frac{1}{V}\int_{A_{12}} w d\mathbf{s} \qquad (10)$$

Last term indicates the interfacial transfer.

Superficial average of time derivatives is

$$\phi \overline{\frac{\partial \mathbf{w}}{\partial t}} = \frac{\partial(\phi \overline{\mathbf{w}})}{\partial t} - \frac{1}{V} \int_{A_{12}} \mathbf{w} \mathbf{u} \cdot d\mathbf{s} \quad (11)$$

The quantities have been defined by volumetric average. However, at the macroscopic boundary the volumetric average fails to apply in region near a macroscopic boundary. It is a serious problem since most important transfer process occurs in boundary layer near wall. To solve this issue, the volumetric average has to be transferred into areal average. The quantities are defined by the area average. At first the mREV is degenerated into a thin plate. Figure 2 shows the schematic method of degenerating an mREV into a very thin layer with the length scale of sREV $dx_3$. Therefore we get a thin plate REV.

$$V = d\overline{x_1} d\overline{x_2} dx_3 \quad (12)$$

For the degenerated REV, $V_2$ is equal to $A_2 dx_3$. So that, we get

$$\phi = \frac{V_2}{V} = \frac{A_2}{A} = \phi_a \quad (13)$$

Therefore, the intrinsic volume average becomes intrinsic areal average.

$$\overline{\mathbf{w}} = \frac{1}{V_2} \int_{V_2} \mathbf{w} dV = \frac{1}{A_2} \int_{A_2} \mathbf{w} dA \quad (14)$$

Superficial areal averages can be obtained for the degenerated REV. The divergent theorem is

$$\phi_a \overline{\nabla \cdot \mathbf{w}} = \overline{\nabla} \cdot (\phi_a \overline{\mathbf{w}}) + \frac{1}{V} \int_{A_{12}} \mathbf{w} \cdot d\mathbf{s} \quad (15)$$

The gradient average is

$$\phi_a \overline{\nabla w} = \overline{\nabla}(\phi_a \overline{w}) + \frac{1}{V} \int_{A_{12}} w d\mathbf{s} \quad (16)$$

The transient average is

$$\phi_a \overline{\frac{\partial w}{\partial t}} = \frac{\partial (\phi_a \overline{w})}{\partial t} - \frac{1}{V} \int_{A_{12}} w \mathbf{u} \cdot d\mathbf{s} \quad (17)$$

The overhead bar is interpreted as the areal averages. The averaged quantities are function of $(\overline{x_1}, \overline{x_2}, x_3)$. As $x_3$ the moves away from the boundary into the core region, areal average recovers the volumetric average.

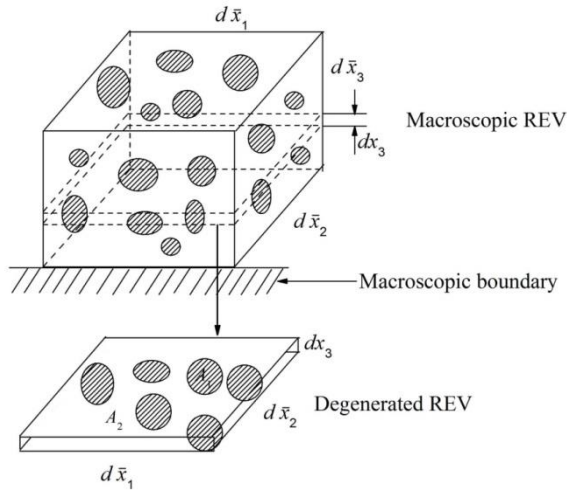

Figure 2 The degeneration of mREV into a thin plate REV

4 Electrochemical double layer and electrochemical reaction occurring at the interface

There are electrochemical double layer and reaction occurring at the interface between the electrolyte fluid and the solid. Therefore, we need to define the physical quantities of the electrochemical double layer and reaction in advance.

The potential of the electrochemical double layer at the interface between the solid and the electrolyte can be calculated by the following equation:

$$\nabla^2 \Phi_d = -\frac{\rho_d}{\varepsilon} = -\frac{\sigma_d}{d} \frac{1}{\varepsilon} \quad (18)$$

where $\Phi_d$ is the potential of the electrochemical double layer, $d$ is the thickness of the double layer, $\rho_d$ and $\sigma_d$ are the volume and the surface charge densities at interface between solid and fluid, respectively.

The electrochemical reaction normally occurs at the interface between the electrode and the electrolyte. One reaction occurring at the interface is

$$O + ne \; \frac{k_f}{k_b} \; R \qquad (19)$$

where $O$ is the reactant, $R$ is the product, $n$ is the charge transfer number of the electrons, and $k_f$ and $k_b$ is the forward and backward reaction rate constants, respectively.

According to the thermodynamics, we get

$$\Delta G = \Delta H - T\Delta S \qquad (20)$$

where $\Delta G$ is the free energy of the reaction per mole (J/mol), $\Delta H$ is enthalpy change per mole (J/mol), and $\Delta S$ is the entropy change per mole.

For $N_O$ mole reactant, the total heat is

$$\text{Total Heat} = (\Delta H - \Delta G)N_O \qquad (21)$$

The rate of heat ($H_E$) per unit area is

$$H_E = (\Delta H - \Delta G)\frac{dN_O}{Sdt} \qquad (22)$$

Therefore, the rate of heat per unit area ($H_E$) is

$$H_E = (\Delta H - \Delta G)\frac{i}{nFS} = \frac{(\Delta H - \Delta G)}{nF}\mathbf{J}_E = \omega \mathbf{J}_E \qquad (23)$$

where $i$ is the current, $S$ is the surface, $\omega$ is coefficient and $\mathbf{J}_E$ is the neat reaction current per unit area.

The neat reaction rate ($v$) and current are

$$v = v_f - v_b = k_f C_O - k_b C_R = \frac{i}{nFS} = \frac{\mathbf{J}_E}{nF} \qquad (24)$$

$$i = i_f - i_b = nFS[k_f C_O - k_b C_R] \quad (25)$$

$$\mathbf{J}_E = \mathbf{J}_f - \mathbf{J}_b = nF[k_f C_O - k_b C_R] \quad (26)$$

where $C_O$ is the concentration of reactant $O$, $C_R$ is the concentration of product $R$, $i_f$ and $i_b$ are the current generated by the forward and backward reactions through a surface of $S$, $\mathbf{J}_f$ and $\mathbf{J}_b$ are current flux generated by the forward and backward reactions through a surface of $S$, respectively. If the reaction is reversible, according to the Bulter-Volmer model, the relation between the current and the overpotential ($\eta$) is

$$i = i_0 [\frac{C_O}{C_0^O} e^{-\frac{nF}{RT}\beta\eta} - \frac{C_R}{C_0^R} e^{-\frac{nF}{RT}(1-\beta)\eta}] \quad (27)$$

$$\mathbf{J}_E = \mathbf{J}_0 [\frac{C_O}{C_0^O} e^{-\frac{nF}{RT}\beta\eta} - \frac{C_R}{C_0^R} e^{-\frac{nF}{RT}(1-\beta)\eta}] \quad (28)$$

where $\beta$ is the transfer constant, $i_0$ is the exchange current, $\mathbf{J}_0$ is the exchange current flux, $C_0^O$ and $C_0^R$ are the bulk concentration in the electrolyte of the reactant $O$ and product $R$, $C_O$ and $C_R$ are the concentration of the reactant $O$ and product $R$, respectively[9].

Note that at the interface we have

$$\mathbf{\Phi}_1 - \mathbf{\Phi}_2 = \eta + \mathbf{\Phi}_d \quad (29)$$

where $\mathbf{\Phi}_1$ and $\mathbf{\Phi}_2$ are potential of solid phase and fluid, respectively. In the practical case, the right has to plus an additive term which depends on the local solution composition[1].

5 Macroscopic governing equations

5.1 Macroscopic governing equations for the integral electrolyte

After the definition of the mREV, there are two REVs in the mixture of the solid and the fluid. One is sREV in the fluid, while the other is the mREV. The governing equations for the microscopic REV have been got in the previous paper. The microscopic governing equations for the integral electrolyte are shown below.

The mass conservation is

$$\nabla \cdot \mathbf{u} = 0 \qquad (30)$$

At the interfacial boundary conditions for rigid media, we get

$$\mathbf{u} = 0 \quad \text{on } A_{12} \qquad (31)$$

The momentum equation is

$$\rho_m \frac{\partial \mathbf{u}}{\partial t} + \rho_m \nabla \cdot (\mathbf{u}\mathbf{u}) = -\nabla p + \mu \nabla \cdot \mathbf{S} \qquad (32)$$

where **S** is strain rate tensor.

The microscopic energy conservation in fluid is

$$\rho_m C_p \frac{\partial T}{\partial t} + \rho_m C_p \nabla \cdot (T\mathbf{u}) = -k\nabla^2 T + H_r \qquad (33a)$$

The microscopic energy conservation in solid is

$$\rho_{m1} C_{p1} \frac{\partial T_1}{\partial t} = -k_1 \nabla^2 T_1 + H_s \qquad (33b)$$

where the subscript 1 represents the solid phase. $H_r$ and $H_s$ are the Joule heat generated inside the REV.

$$H_r = \sigma_2 \nabla \Phi_2 \cdot \nabla \Phi_2 \qquad (34a)$$

$$H_s = \sigma_1 \nabla \Phi_1 \cdot \nabla \Phi_1 \qquad (34b)$$

where $\sigma_1$ and $\sigma_2$ are the conductivity of solid and electrolyte, respectively.

At the interfacial, there are boundary conditions:

$$T = T_1 \, ; \, \boldsymbol{n} \cdot k\nabla T = \boldsymbol{n} \cdot k\nabla T_1 + \boldsymbol{n} \cdot H_E \text{ on } A_{12} \qquad (35)$$

where the subscript 1 means the solid phase and $\boldsymbol{n}$ is the unit vector out normal from fluid to solid. Note that the heat flux at the interface will be influenced by the heat generated at the interface by the electrochemical reaction.

After the definition of microscopic equations, we have to obtain the macroscopic equations. At first, the velocity and pressure are decomposed into

$$\mathbf{u} = \bar{\mathbf{u}} + \mathbf{u}' \quad (36a)$$

$$p = \bar{p} + p' \quad (36b)$$

$$T = \bar{T} + T' \quad (36c)$$

$$C_i = \bar{C}_i + C_i' \quad (36d)$$

where the overhead bar means the average quantities and the last term with superscript is the dispersion of the quantities.

Then the macroscopic equations can be obtained by performing the intrinsic volumetric averages to the microscopic equations.

The macroscopic mass conservation is

$$\phi \overline{\nabla \cdot \mathbf{u}} = \phi \overline{\nabla \cdot (\bar{\mathbf{u}} + \mathbf{u}')} = \bar{\nabla} \cdot (\phi \bar{\mathbf{u}} + \phi \bar{\mathbf{u}}') = \bar{\nabla} \cdot (\phi \bar{\mathbf{u}}) = 0 \quad (37)$$

The macroscopic momentum conservations is

$$\rho_m \left[ \frac{\partial}{\partial t} (\phi \bar{\mathbf{u}}) + \bar{\nabla} \cdot (\phi \bar{\mathbf{u}} \bar{\mathbf{u}}) \right] = -\bar{\nabla}(\phi \bar{p}) + \rho_m [\nu \bar{\nabla} \cdot (\phi \bar{\mathbf{S}}) + \bar{\nabla} \cdot (-\phi \overline{\mathbf{u}'\mathbf{u}'})] + \bar{\mathbf{b}} \quad (38)$$

There are new unknown terms in momentum equations. The first is the momentum dispersion of $-\phi \overline{\mathbf{u}'\mathbf{u}'}$. And the interfacial force between the solid phase and the fluid phase.

$$\bar{\mathbf{b}} = \frac{1}{V} \int_{A_{12}} (-p\mathbf{I} + \mu \mathbf{S}) \cdot d\mathbf{s} = \frac{1}{V} \int_{A_{12}} (-p'\mathbf{I} + \mu \mathbf{S}') \cdot d\mathbf{s} \quad (39)$$

The macroscopic energy conservation in fluid is

$$\rho_m C_p \frac{\partial (\phi \bar{T})}{\partial t} + \rho_m C_p \bar{\nabla} \cdot (\bar{T}\bar{\mathbf{u}}) = -k\bar{\nabla}^2(\phi\bar{T}) + \rho_m C_p \bar{\nabla} \cdot (-\phi\overline{T'\mathbf{u}'})$$

$$+ k\bar{\nabla} \cdot \Lambda_{12} + \mathbf{q}_{12} + \frac{\sigma_2}{\phi}\bar{\nabla}(\phi\bar{\Phi}_2) \cdot \bar{\nabla}(\phi\bar{\Phi}_2) + \sigma_2 \phi \overline{\nabla\Phi'_2 \cdot \nabla\Phi'_2} \quad (40)$$

The macroscopic energy conservation in solid is

$$\rho_{m1} C_p \frac{\partial (\phi_1 \bar{T}_1)}{\partial t} = -k_1 \bar{\nabla}^2(\phi_1 \bar{T}_1) - k_1 \bar{\nabla} \cdot \Lambda_{12} - \mathbf{q}_{12}$$

$$+ \frac{\sigma_1}{\phi_1}\bar{\nabla}(\phi_1\bar{\Phi}_1) \cdot \bar{\nabla}(\phi_1\bar{\Phi}_1) + \sigma_1 \phi_1 \overline{\nabla\Phi'_1 \cdot \nabla\Phi'_1} \quad (41)$$

Note that the sum of porosity of fluid and solid is equal to 1.

$$\phi_1 + \phi = 1 \quad (42)$$

There are new unknown terms in the momentum conservation equation. The energy dispersion is $-\phi\overline{T'\mathbf{u}'}$. The Joule heat dispersion are $\phi\overline{\nabla\Phi'_2 \cdot \nabla\Phi'_2}$ and $\phi_1\overline{\nabla\Phi'_1 \cdot \nabla\Phi'_1}$ in the fluid and the solid, respectively.

Thermal tortuosity is

$$\Lambda_{12} = \frac{1}{V}\int_{A_{12}} T \cdot d\mathbf{s} = \frac{1}{V}\int_{A_{12}} T' \cdot d\mathbf{s} \quad (43)$$

The interfacial heat transfer is

$$\mathbf{q}_{12} = \frac{1}{V}\int_{A_{12}} k\nabla T \cdot d\mathbf{s} \quad (44)$$

The macroscopic governing equations for the electrolyte are summarized in Table 2. The new unknown terms are listed in Table 3.

Table 2 The macroscopic governing equations of the integral electrolyte

| Name | | Governing equation |
|---|---|---|
| Mass conservation | | $\overline{\nabla} \cdot (\phi \overline{\mathbf{u}}) = 0$ |
| Momentum conservation | | $\rho_m \left[ \dfrac{\partial}{\partial t}(\phi \overline{\mathbf{u}}) + \overline{\nabla} \cdot (\phi \overline{\mathbf{u}}\overline{\mathbf{u}}) \right] = -\overline{\nabla}(\phi \bar{p}) + \rho_m [\nu \overline{\nabla} \cdot (\phi \overline{\mathbf{S}}) + \overline{\nabla} \cdot (-\phi \overline{\mathbf{u}'\mathbf{u}'})] + \bar{\mathbf{b}}$ |
| Energy equation | Fluid | $\rho_m C_p \dfrac{\partial(\phi \bar{T})}{\partial t} + \rho_m C_p \overline{\nabla} \cdot (\bar{T}\bar{u})$ $= -k\overline{\nabla}^2(\phi \bar{T}) + \rho_m C_p \overline{\nabla} \cdot (-\phi \overline{T'u'}) + k\overline{\nabla} \cdot \Lambda_{12} + \mathbf{q}_{12}$ $+ \dfrac{\sigma_2}{\phi} \overline{\nabla}(\phi \overline{\mathbf{\Phi}}_2) \cdot \overline{\nabla}(\phi \overline{\mathbf{\Phi}}_2) + \sigma_2 \phi \overline{\nabla \mathbf{\Phi}'_2 \cdot \nabla \mathbf{\Phi}'_2}$ |
| | Solid | $\rho_{m1} C_p \dfrac{\partial(\phi_1 \bar{T}_1)}{\partial t} = -k_1 \overline{\nabla}^2(\phi_1 \bar{T}_1) - k_1 \overline{\nabla} \cdot \Lambda_{12} - \mathbf{q}_{12}$ $+ \dfrac{\sigma_1}{\phi_1} \overline{\nabla}(\phi_1 \overline{\mathbf{\Phi}}_1) \cdot \overline{\nabla}(\phi_1 \overline{\mathbf{\Phi}}_1) + \sigma_1 \phi_1 \overline{\nabla \mathbf{\Phi}'_1 \cdot \nabla \mathbf{\Phi}'_1}$ |

Table 3 The new unknown terms in the governing equations for the integral electrolyte

| Governing equation | Name | New unknown term |
|---|---|---|
| Momentum conservation | Momentum dispersion | $-\phi \overline{\mathbf{u}'\mathbf{u}'}$ |
| | Interfacial force | $\bar{\mathbf{b}} = \dfrac{1}{V} \int_{A_{12}} (-p\mathbf{I} + \mu \mathbf{S}) \cdot d\mathbf{s}$ |
| Energy equation | Energy dispersion | $-\phi \overline{T'\mathbf{u}'}$ |

|   |   |   | $\phi \overline{\nabla \Phi_2' \cdot \nabla \Phi_2'}$ |
|---|---|---|---|
|   |   |   | $\phi_1 \overline{\nabla \Phi_1' \cdot \nabla \Phi_1'}$ |
|   | Thermal tortuosity |   | $\Lambda_{12} = \dfrac{1}{V}\int_{A_{12}} T \cdot d\mathbf{s} = \dfrac{1}{V}\int_{A_{12}} T' \cdot d\mathbf{s}$ |
|   | Interfacial heat transfer |   | $\mathbf{q}_{12} = \dfrac{1}{V}\int_{A_{12}} k\nabla T \cdot d\mathbf{s}$ |

5.2 Macroscopic governing equation for species *i*

5.2.1 Charge conservation

In the bulk solid or electrolyte, it is electro-neutral. Therefore we get the microscopic equations for solid and the electrolyte

$$\nabla^2 \Phi_1 = 0 \quad (45a)$$

$$\nabla^2 \Phi_2 = 0 \quad (45b)$$

where $\Phi_1$ and $\Phi_2$ are the potentials of the solid and the fluid, respectively.

Therefore the macroscopic governing equations for the solid and the fluid is

$$\overline{\nabla}^2 (\phi_1 \overline{\Phi}_1) - \Xi_1 - \Psi_1 = 0 \qquad (46a)$$

$$\overline{\nabla}^2 (\phi \overline{\Phi}_2) + \Xi_2 + \Psi_2 = 0 \qquad (46b)$$

At the interface between the solid and the electrolyte fluid, there is the electrochemical double layer. Therefore, the macroscopic charge conservation at interface becomes:

$$\overline{\nabla}^2(\phi\overline{\Phi}_d) = -\frac{1}{\varepsilon}\sum_i^n z_i F \frac{1}{V}\int_{A_{12}} C_i \cdot d\mathbf{s} = -\frac{1}{\varepsilon}\sum_i^n z_i F \Omega_i \quad (47)$$

where $\overline{\Phi}_d$ is the average potential of the electrochemical double layer.

$\Xi_1$ and $\Xi_2$ are tortuosity of electric fields in the solid and the fluid, respectively. $\Psi_1$ and $\Psi_2$ are tortuosity of potentials in the solid and the fluid, respectively.

$$\Xi_1 = \frac{1}{V}\int_{A_{12}} \nabla\Phi_1 \cdot d\mathbf{s} \quad (48a)$$

$$\Psi_1 = \frac{1}{V}\int_{A_{12}} \Phi_1 \cdot d\mathbf{s} \quad (48b)$$

$$\Xi_2 = \frac{1}{V}\int_{A_{12}} \nabla\Phi_2 \cdot d\mathbf{s} \quad (48c)$$

$$\Psi_2 = \frac{1}{V}\int_{A_{12}} \Phi_2 \cdot d\mathbf{s} \quad (48d)$$

$\Omega_i$ is the mass transfer tortuosity of species $i$ at the interface.

$$\Omega_i = \frac{1}{V}\int_{A_{12}} C_i \cdot d\mathbf{s} = \frac{1}{V}\int_{A_{12}} C_i' \cdot d\mathbf{s} \quad (49)$$

5.2.2 Mass conservation

The mass conservation for macroscopic REV is

$$\frac{\partial(\phi\bar{C}_i)}{\partial t} + \overline{\nabla} \cdot (\phi\bar{C}_i\overline{\mathbf{u}}_i) = \overline{\nabla} \cdot (-\phi\overline{C_i'\mathbf{u}_i'}) + \dot{m}_{ci} \quad (50)$$

5.2.3 The microscopic momentum equation is

$$-\nabla p_i + \mu\nabla^2 \mathbf{u}_i - z_i F C_i \nabla\Phi_2 = 0 \quad (51)$$

The macroscopic equation of momentum conservation is

$$-\overline{\nabla}(\phi \bar{p}_i) + \mu \overline{\nabla} \cdot (\phi \bar{\mathbf{S}}_i) + \bar{\mathbf{b}}_i - z_i F \bar{C}_i \overline{\nabla}(\phi \overline{\Phi}_2) - z_i F \phi \overline{C'_i \nabla \Phi'_2} - z_i F \mathbf{\Gamma}_i = 0 \quad (52)$$

where $\bar{\mathbf{S}}_i$ is the stress rate tensor of ion species $i$, $\bar{\mathbf{b}}_i$ is the interfacial force of ion species $i$, $\mathbf{\Gamma}_i$ is the tortuosity of electric force and $z_i F \phi \overline{C'_i \nabla \Phi'_2}$ is the dispersion of electric force.

$$\bar{\mathbf{b}}_i = \frac{1}{V} \int_{A_{12}} (-p_i \mathbf{I} + \mu \mathbf{S}_i) \cdot d\mathbf{s} \quad (53)$$

$$\mathbf{\Gamma}_i = \frac{1}{V} \int_{A_{12}} C_i \nabla \Phi_2 \cdot d\mathbf{s} \quad (54)$$

5.2.4 Concentration conservation

In the REV the microscopic concentration equations in the fluid and the solid can be rewritten as

$$\frac{\partial C_i}{\partial t} = D_i \nabla^2 C_i + \nabla \cdot (\gamma C_i \nabla \Phi_2) - \nabla \cdot (C_i \mathbf{u}_i) \quad (55)$$

$$\frac{\partial C_{1i}}{\partial t} = D_{1i} \nabla^2 C_{1i} + \gamma_1 \nabla \cdot (C_{1i} \nabla \Phi_1) \quad (56)$$

where $\gamma$ and $\gamma_1$ are coefficients.

$$\gamma = \frac{z_i F D_i}{RT} \quad (57)$$

$$\gamma_1 = \frac{z_i F D_{1i}}{RT} \quad (58)$$

where $D_i$ and $D_{1i}$ are the diffusion coefficients of species $i$ in the fluid and the solid phase, respectively. Note that we delete the mass source term because it can be easily added. The proper boundary conditions on the fluid-solid interface of $A_{12}$ are

$$C_{1i} = C_i \quad on\ A_{12} \quad (59a)$$

$$\mathbf{n} \cdot D_{1i} \nabla C_{1i} = \mathbf{n} \cdot D_i \nabla C_i \qquad \text{on } A_{12} \qquad (59b)$$

and

$$\mathbf{\Phi}_1 = \mathbf{\Phi}_2 + \eta + \mathbf{\Phi}_d \qquad \text{on } A_{12} \qquad (59c)$$

$$\mathbf{n} \cdot \gamma_1 C_{1i} \nabla \mathbf{\Phi}_1 = \mathbf{n} \cdot \gamma C_i \nabla (\mathbf{\Phi}_2 + \eta + \mathbf{\Phi}_d) \qquad \text{on } A_{12} \qquad (59d)$$

The macroscopic concentration conservation is

$$\frac{\partial (\phi \bar{C}_i)}{\partial t} + \bar{\nabla} \cdot (\phi \bar{C}_i \bar{\mathbf{u}}_i) = D_i \bar{\nabla}^2 (\phi \bar{C}_i) + D_i \bar{\nabla} \cdot \mathbf{\Omega}_i + m_{12} + \gamma \bar{\nabla} \cdot [\bar{C}_i \bar{\nabla}(\phi \bar{\mathbf{\Phi}}_2)]$$

$$+ \gamma \bar{\nabla} \cdot \mathbf{\Gamma}_i + \gamma \mathbf{\Gamma}_i - \bar{\nabla} \cdot (-\phi \gamma \overline{C'_i \nabla \mathbf{\Phi}'_2}) + \bar{\nabla} \cdot (-\phi \overline{C'_i \mathbf{u}'_i}) \qquad (60)$$

There are unknown terms in the equation. Mass dispersion is $-\phi \overline{C'_i \mathbf{u}'_i}$.
Interfacial mass transfer from solid into fluid is

$$m_{12} = \frac{1}{V} \int_{A_{12}} D_i \nabla C_i \cdot d\mathbf{s} \qquad (61)$$

The macroscopic equation for the transport of ion species in the solid is

$$\frac{\partial (\phi_1 \bar{C}_{1i})}{\partial t} = D_{1i} \bar{\nabla}^2 (\phi_1 \bar{C}_{1i}) + \gamma_1 \bar{\nabla} \cdot [\bar{C}_{1i} \bar{\nabla}(\phi_1 \bar{\mathbf{\Phi}}_1)] - \gamma_1 \bar{\nabla} \cdot \mathbf{\Gamma}_i - \gamma_1 \mathbf{\Gamma}_i$$

$$- \bar{\nabla} \cdot (-\phi_1 \gamma_1 \overline{C'_{1i} \nabla \mathbf{\Phi}'_1}) - D_{1i} \bar{\nabla} \cdot \mathbf{\Omega}_i - m_{12} + m_{ad} \qquad (62)$$

where $C_{1i}$ is the concentration of ion species $i$ in solid phase and $m_{ad}$ is the interfacial mass adsorption on the surface of the solid.

$$m_{ad} = \frac{1}{V} \int_{A_{12}} \mathbf{m}''_{ad} \cdot d\mathbf{s} \qquad (63)$$

$D_i$ and $D_{1i}$ are the diffusion diffusivity coefficients of ion species $i$ in the fluid and solid, respectively.

5.2.5 Current equations

For the macroscopic REV, the current comes from three contributions including solid, interface and fluid.

In the fluid, the total current is

$$\mathbf{J} = \sum \mathbf{J}_i \qquad (64)$$

$\mathbf{J}_i$ is the current flux generated by the motion of ion species $i$.

$$\mathbf{J}_i = -z_i F[D_i \nabla C_i + \gamma C_i \nabla \Phi_2 - C_i \mathbf{u}_i] \qquad (65)$$

The macroscopic current flux from the motion of the ions in the fluid phase is

$$\phi \bar{\mathbf{J}}_2 = \sum \phi \bar{\mathbf{J}}_i \qquad (66a)$$

$$\phi \bar{\mathbf{J}}_i = -z_i F[D_i \bar{\nabla}(\phi \bar{C}_i) + D_i \boldsymbol{\Omega}_i + \gamma \bar{C}_i \bar{\nabla}(\phi \bar{\Phi}_2) + \gamma \phi \overline{C'_i \nabla \Phi'_2} + \gamma \boldsymbol{\Gamma}_i - \bar{C}_i \bar{\mathbf{u}}_i + (-\phi \overline{C'_i \mathbf{u}'_i})] \qquad (66b)$$

The total current flux ($\mathbf{J}_s$) in solid phase generated by the motion of ions and electron is

$$\mathbf{J}_s = \mathbf{J}_1 + \mathbf{J}_e \qquad (67)$$

where $\mathbf{J}_1$ and $\mathbf{J}_e$ are current flux of ions and electron in the solid phase, respectively.

The current flux in the solid phase generated by the motion of ions is

$$\mathbf{J}_{1i} = -z_i F[D_{1i} \nabla C_{1i} - \gamma_1 C_{1i} \nabla \Phi_1] \qquad (68)$$

Similarly, the current flux in the solid phase is

$$\phi_1 \bar{\mathbf{J}}_1 = \sum \phi_1 \bar{\mathbf{J}}_{1i} \qquad (69a)$$

$$\phi_1 \bar{\mathbf{J}}_{1i} = -z_i F [D_{1i} \bar{\nabla}(\phi_1 \bar{C}_{1i}) - D_{1i} \Omega_i + \gamma_1 \bar{C}_{1i} \bar{\nabla}(\phi_1 \bar{\Phi}_1) + \gamma_1 \phi \overline{C'_{1i} \nabla \Phi'_1} - \gamma_1 \Gamma_i] \quad (69b)$$

The movement of electrons in the solid is governed by Ohm's law.

$$\mathbf{J}_e = -\sigma_1 \nabla \Phi_1 \quad (70)$$

Note that from the statics point of view, Ohm's law is useful to descript the current flux generated by the movement of electrons [1,2].

The macroscopic current flux of electrons in solid is

$$\phi_1 \bar{\mathbf{J}}_e = -\sigma_1 \bar{\nabla}(\phi_1 \bar{\Phi}_1) - \sigma_1 \Psi_1 \quad (71)$$

The current flux due to the electrochemical reaction is in general form

$$\mathbf{J}_E = nF[k_f C_O - k_b C_R] \quad (72a)$$

For the reversible reaction, Bulter-Volmer equation says

$$\mathbf{J}_E = \mathbf{J}_0 \left[ \frac{C_O}{C_0^O} e^{-\frac{nF}{RT}\beta\eta} - \frac{C_R}{C_0^R} e^{-\frac{nF}{RT}(1-\beta)\eta} \right] \quad (72b)$$

The macroscopic current flux generated at the interface is

$$\phi \bar{\mathbf{J}}_E = nF[k_f \phi \bar{C}_O - k_b \phi \bar{C}_R] \quad (73a)$$

or

$$\phi \bar{\mathbf{J}}_E = \mathbf{J}_0 \left[ \frac{\phi \bar{C}_O}{C_0^O} e^{-\frac{nF}{R\bar{T}}\beta\bar{\eta}} - \frac{\phi \bar{C}_R}{C_0^R} e^{-\frac{nF}{R\bar{T}}(1-\beta)\bar{\eta}} \right] \quad (73b)$$

The macroscopic governing equations for ion species $i$ in the fluid and solid are summarized in the Table 4. There are new unknown terms appearing in the governing equations. Table 8 summarizes the new unknown terms.

Table 4 Summary of the macroscopic governing equations for the species $i$

| Name | | Equation |
|---|---|---|
| Charge conservation | | $\overline{\nabla}^2(\phi_1\overline{\boldsymbol{\Phi}}_1) - \Xi_1 - \Psi_1 = 0$ <br><br> $\overline{\nabla}^2(\phi\overline{\boldsymbol{\Phi}}_2) + \Xi_2 + \Psi_2 = 0$ <br><br> $\overline{\nabla}^2(\phi\overline{\boldsymbol{\Phi}}_d) = -\dfrac{1}{\varepsilon}\sum_{i}^{n} z_i F \Omega_i$ |
| Mass conservation | | $\dfrac{\partial(\phi\bar{C}_i)}{\partial t} + \overline{\nabla}\cdot(\phi\bar{C}_i\overline{\mathbf{u}}_i) = \overline{\nabla}\cdot(-\phi\overline{C'_i\mathbf{u}'_i}) + \dot{m}_{ci}$ |
| Momentum conservation | | $-\overline{\nabla}(\phi\bar{p}) + \mu\overline{\nabla}\cdot(\phi\overline{\mathbf{S}}_i) + \overline{\mathbf{b}}_i - z_i F \bar{C}_i \overline{\nabla}(\phi\overline{\boldsymbol{\Phi}}_2) - z_i F \phi \overline{C'_i \nabla\boldsymbol{\Phi}'_2} - z_i F \boldsymbol{\Gamma}_i = 0$ |
| Concentration conservation | Flow | $\dfrac{\partial(\phi\bar{C}_i)}{\partial t} + \nabla\cdot(\phi\bar{C}_i\overline{\mathbf{u}}_i)$ <br><br> $= D_i\overline{\nabla}^2(\phi\bar{C}_i) + D_i\overline{\nabla}\cdot\boldsymbol{\Omega}_i + m_{12} + \gamma\overline{\nabla}\cdot[\bar{C}_i\overline{\nabla}(\phi\overline{\boldsymbol{\Phi}}_2)]$ <br><br> $+\gamma\overline{\nabla}\cdot\boldsymbol{\Gamma}_i + \gamma\boldsymbol{\Gamma}_i - \overline{\nabla}\cdot(-\phi\gamma\overline{C'_i\nabla\boldsymbol{\Phi}'_2}) + \overline{\nabla}\cdot(-\phi\overline{C'_i\mathbf{u}'_i})$ |
| | Solid | $\dfrac{\partial(\phi_1\bar{C}_{1i})}{\partial t} = D_{1i}\overline{\nabla}^2(\phi_1\bar{C}_{1i}) + \gamma_1\overline{\nabla}\cdot[\bar{C}_{1i}\overline{\nabla}(\phi\overline{\boldsymbol{\Phi}}_1)] - \gamma_1\overline{\nabla}\cdot\boldsymbol{\Gamma}_i - \gamma_1\boldsymbol{\Gamma}_i$ <br><br> $-\overline{\nabla}\cdot(-\phi_1\gamma_1\overline{C'_{1i}\nabla\boldsymbol{\Phi}'_1}) - D_{1i}\overline{\nabla}\cdot\boldsymbol{\Omega}_i - m_{12} + m_{ad}$ |
| Current equation | Flow | $\phi\overline{\mathbf{J}}_i = -z_i F[D_i\overline{\nabla}(\phi\bar{C}_i) + D_i\boldsymbol{\Omega}_i + \gamma\bar{C}_i\overline{\nabla}(\phi\overline{\boldsymbol{\Phi}}_2) + \gamma\phi\overline{C'_i\nabla\boldsymbol{\Phi}'_2} + \gamma\boldsymbol{\Gamma}_i - \bar{C}_i\overline{\mathbf{u}}_i$ <br><br> $+ (-\phi\overline{C'_i\mathbf{u}'_i})]$ |
| | Solid | $\phi_1\overline{\mathbf{J}}_{1i} = -z_i F[D_{1i}\overline{\nabla}(\phi_1\bar{C}_{1i}) - D_{1i}\boldsymbol{\Omega}_i + \gamma_1\bar{C}_{1i}\overline{\nabla}(\phi_1\overline{\boldsymbol{\Phi}}_1) + \gamma_1\phi\overline{C'_{1i}\nabla\boldsymbol{\Phi}'_1}$ <br><br> $- \gamma_1\boldsymbol{\Gamma}_i]$ |

|  |  | $\phi_1 \bar{\mathbf{J}}_e = -\sigma_1 \bar{\nabla}(\phi_1 \bar{\mathbf{\Phi}}_1) - \sigma_1 \mathbf{\Psi}_1$ |
|---|---|---|
|  | **Interface** | $\phi \bar{\mathbf{J}}_E = nF[k_f \phi \bar{C}_O - k_b \phi \bar{C}_R]$ |

Table 5 The new unknown terms in the governing equation for species $i$

| Governing equation | Name | New unknown term |
|---|---|---|
| Charge conservation | Tortuosity of electric field | $\mathbf{\Xi}_1 = \frac{1}{V} \int_{A_{12}} \nabla \mathbf{\Phi}_1 \cdot d\mathbf{s}$ |
|  |  | $\mathbf{\Xi}_2 = \frac{1}{V} \int_{A_{12}} \nabla \mathbf{\Phi}_2 \cdot d\mathbf{s}$ |
|  | Tortuosity of potential | $\mathbf{\Psi}_1 = \frac{1}{V} \int_{A_{12}} \mathbf{\Phi}_1 \cdot d\mathbf{s}$ |
|  |  | $\mathbf{\Psi}_2 = \frac{1}{V} \int_{A_{12}} \mathbf{\Phi}_2 \cdot d\mathbf{s}$ |
| Momentum equation | Interfacial force | $\bar{\mathbf{b}}_i = \frac{1}{V} \int_{A_{12}} (-p_i \mathbf{I} + \mu \mathbf{S}_i) \cdot d\mathbf{s}$ |
|  | Tortuosity of electric force | $\mathbf{\Gamma}_i = \frac{1}{V} \int_{A_{12}} C_i \nabla \mathbf{\Phi}_2 \cdot d\mathbf{s}$ |
| | Dispersion of electric force | $-\phi\overline{C'\nabla\Phi'}$ |
| Concentration conservation | Mass dispersion | $-\phi\overline{C_i'\mathbf{u}_i'}$ |
| | Mass tortuosity | $\mathbf{\Omega}_i = \dfrac{1}{V}\int_{A_{12}} C_i \cdot d\mathbf{s} = \dfrac{1}{V}\int_{A_{12}} C_i' \cdot d\mathbf{s}$ |
| | Interfacial mass transfer | $m_{12} = \dfrac{1}{V}\int_{A_{12}} D_i \nabla C_i \cdot d\mathbf{s}$ |
| | Interfacial mass adsorption. | $m_{ad} = \dfrac{1}{V}\int_{A_{12}} \mathbf{m}_{ad}''$ |

6. Concluding remarks

The fluid (liquid or gas) flow-through the porous media is frequently encountered in nature or in many engineering applications, such as the electrochemical systems, manufacturing of composites, oil production, geothermal engineering, nuclear thermal disposal, soil pollution, to name a few. Electrochemical systems, for example, electrochemical reactors, batteries, supercapacitors, fuel cells, etc, are widely used in modern society. The common and fundamental process involving in the electrochemical systems is the electrolyte flows through the porous electrode. In this study, we establish the general theory for the electrochemical flows-through porous media. We use static method and set up two representative elementary volumes (REVs). One is the macroscopic REV of the mixture of the porous media and the electrolyte (or gas), while the other is the microscopic REV in the electrolyte fluid. The establishment of two REVs enables us to investigate the details of transports of mass, heat, electric flied, or momentum in the process of the electrochemical flows-through porous electrode. In our previous paper, we derived the governing equations from the conservation laws of charge, mass, momentum, energy and mass concentration in the microscopic REV. In addition, we normalize the governing equations to derive the dimensionless parameters, known as Reynolds, Thompson, Peclet, Prandtl and *X* numbers. A new number, named *X* number, appears in the Navier-Stokes equation symbolizing the balance between the inertia force and the electric force.

In this work, the macroscopic governing equations are derived from the conservation laws in the macroscopic REV, the mixture of the solid and the fluid. At first, we define the porosity by the volume and surface and divide the porosity into various categories. Then the superficial averages is transformed into intrinsic averages to derive the interaction terms between the solid and the fluid, known as dispersion, tortuosity and interfacial transfer. The macroscopic governing equations are derived by performing the intrinsic average on the microscopic governing equations. After done that, the unknown terms related to the dispersion, tortuosity and interfacial transfer are emerged in the governing equations. Our new theory provides a new approach to the electrochemical flows-through porous media. It is suitable for the general case, in which the electrolyte fluid flows through the porous media. In our next work, the closure models will be established to derive the expression of the unknown terms of the dispersion, tortuosity and interfacial transfer.

**Acknowledgements**

The author C. Xu would like to thank Dr. Sum Wai Chiang for helpful discussion. We like to thank the financial support from National Nature Science Foundation of China under Grants (No. 51102139) and from Shenzhen Technical Plan Projects (No. JC201105201100A). We also thank the financial support from Guangdong Province Innovation R&D Team Plan (2009010025). CT would like to thank City Key Laboratory of Thermal Management Engineering and Materials for financial support.

**Appendix**

Notation

REV representative elementary volume

mREV macroscopic representative elementary volume

sREV microscopic representative elementary volume

**w** physical quantities used for macroscopic representative elementary volume

$\overline{\mathbf{w}}$ average of a fluid quantity **w** over the fluid volume

W average of a fluid quantity **w** over total volume

V volume of macroscopic representative

elementary volume

$V_1$ volume of solid phase in macroscopic representative elementary volume

$V_2$ volume of fluid phase in macroscopic representative elementary volume

**r** location

$t$ time

$\nabla$ Hamilton operator

$\nabla^2$ Laplacian operator

$\overline{\nabla}$ macroscopic gradient operator

$L$ length scale of the solid

$l$ length scale of the macroscopic REV

$d\overline{x_1}$ scale length of macroscopic REV

$d_p$ particle size

$l_m$ scale length of molecule

$dx_i$ scale length of microscopic REV

$\rho_m$ mass density of fluid

$\rho_c$ total charge density

$z_i$ charge number of species $i$

$n$ numbers of ion species

$n_i$ ion numbers of ion species $i$

$\rho_{mi}$ mass density of species $i$

$m_i$ mass of one particle of species $i$

$C_i$ concentration of species $i$ in the fluid

$C_i^0$ concentration of species $i$ in the bulk electrolyte

$C_{1i}$ concentration of species $i$ in the solid

$M_i$ atomic weight of species $i$

$F$ Faradic constant

$\rho_{ci}$ charge density of species $i$

$x_1$ one axis of the Cartesian coordinate

$x_2$ one axis of the Cartesian coordinate

$x_3$ one axis of the Cartesian coordinate

**u** mean velocity of the fluid

$\overline{\mathbf{u}}$ intrinsic velocity of the fluid

$\mathbf{u}'$ dispersion of the velocity of the fluid

$\mathbf{u}_s$ mean velocity of the solvent group

$\mathbf{u}_i$ mean velocity of species $i$

$\mathbf{v}_i'$ thermal velocity of species $i$

$T$ temperature

$S$ surface area

$s$ unit surface

$\mathbf{n}$ unite surface vector

$D_i$ mass diffusion coefficient of species $i$ in the fluid

$D_{1i}$ diffusion coefficient of species $i$ in the solid phase

**S** strain rate tensor

$C_p$ thermal capacity of the fluid

$k$ thermal conductivity of the fluid

$k_1$ thermal conductivity of the solid

$p$ pressure

$T_2$ temperature of the electrochemical fluid

$T_1$ temperature of solid phase

$A$ overall area of the surface of macroscopic representative elementary volume

$A_1$ solid-solid area on the surface of macroscopic representative elementary volume

$A_2$ fluid-fluid area on the surface of macroscopic representative elementary volume

$A_{12}$ fluid-solid interface area in macroscopic representative elementary volume

$O$ reactant in electrochemical reaction

$R$ product in electrochemical reaction

$C_O$ concentration of reactant $O$

$C_R$ concentration of product $R$

$C_0^O$ bulk concentration in the electrolyte of the reactant $O$

$C_0^R$ bulk concentration in the electrolyte of the product $R$

$n$ charge transfer number of the electrons in electrochemical reaction

$k_f$ forward reaction rate constant in electrochemical reaction

$k_b$ backward reaction rate constant in electrochemical reaction

$\Delta G$ free energy of the reaction per mole in electrochemical reaction

$\Delta H$ enthalpy change per mole in electrochemical reaction

$\Delta S$ entropy change per mole in electrochemical reaction

$H_E$ rate of heat per unit area generated by electrochemical reaction

$H_s$ heat generated in the solid phase

$H_r$ heat generated inside the fluid

$i$ current

$i_f$ current generated by the forward reaction through a surface of $S$

$i_b$ current generated by the backward reaction through a surface of $S$

$i_0$ exchange current

$\mathbf{J}_f$ current flux generated by the forward reaction through a surface of $S$

$\mathbf{J}_b$ current flux generated by the backward reaction through a surface of $S$

$\mathbf{J}_i$ current flux of species $i$ in the fluid

$\mathbf{j}_i$ current flux of species $i$ due to mass diffusion

$\mathbf{J}_0$ exchange current flux

$\mathbf{J}$ current flux

$\mathbf{J}_E$ neat reaction current per unit area

$\mathbf{J}_e$ current flux of electron in the solid phase

$\mathbf{J}_1$ current flux of ions in the solid phase

$\mathbf{J}_2$ current flux of ions in the fluid phase

$\mathbf{J}_s$ total current flux in the solid phase

$\overline{\mathbf{u}'\mathbf{u}'}$ momentum dispersion

$\overline{\mathbf{b}}$ interfacial force between the solid phase and the fluid phase of solvent

$\overline{T'\mathbf{u}'}$ energy dispersion

$\Lambda_{12}$ thermal tortuosity

$\mathbf{q}_{12}$ interfacial heat transfer

$\overline{C_i'\mathbf{u}_i'}$ mass dispersion

$\Xi_1$ tortuosity of electric field in the solid

$\Xi_2$ tortuosity of electric field in the fluid

$\Psi_1$ tortuosity of potential in the solid

$\Psi_2$ tortuosity of potential in the fluid

$\Omega_i$ mass transfer tortuosity of species $i$ at the interface

$m_{12}$ interfacial mass transfer from solid into fluid

$m_{ad}$ interfacial mass adsorption on the surface

of the solid

$\overline{\mathbf{S}}_i$ stress rate tensor of species $i$

$\overline{\mathbf{b}}_i$ interfacial force of species $i$

$\mathbf{\Gamma}_i$ tortuosity of electric force of species $i$

$\overline{C'\nabla\Phi'}$ dispersion of electric force

Greek letters

$\Bbbk$ Boltzmann constant

$\omega$ constant in heat generated by the electrochemical reaction

$\varepsilon$ dielectric constant of the electrolyte

$\lambda_i$ diffusion conductivity

$\tau$ time scale of one electrochemical fluid

$\mathbf{\Phi}_1$ potential of solid phase and fluid

$\mathbf{\Phi}_2$ potential of electrochemical fluid

$\mathbf{\Phi}_d$ potential of the electrochemical double layer

$\eta$ overpotential of electrochemical reaction

$\mu$ dynamic viscosity of the electrolyte

$\nu$ kinematic viscosity of the electrolyte

$\alpha$ thermal diffusivity of the fluid

$\phi$ volume porosity of the fluid

$\phi_a$ area porosity of the fluid

$\phi_1$ volume porosity of the solid

$\sigma_1$ conductivity of the solid phase

$\sigma_2$ conductivity of the fluid phase

$\gamma$ coefficients of the fluid in mass concentration conservation equation

$\gamma_1$ coefficient of the solid in mass concentration conservation equation

Subscripts

1 solid phase

2 fluid phase

$i$ species

Superscripts

$'$ dispersion

Overhead bar

$-$ intrinsic average of physical quantity